\documentclass[preprint2]{aastex}
\usepackage{multirow}

\slugcomment{Chinese Spectral Radioheliograph}

\shorttitle{Distributed Data Processing Pipeline For Chinese Spectral Radioheliograph}
\shortauthors{F. Wang et al.}

\begin{document}


\title{Distributed Data Processing Pipeline For Chinese Spectral Radioheliograph}


\author{F. Wang\altaffilmark{1, 2, 3} Y. Mei H. Deng\altaffilmark{1} C.Y. Liu\altaffilmark{1} D.H. Liu\altaffilmark{4}  S.L. Wei\altaffilmark{1, 2, 3} W. Dai\altaffilmark{1, 2, 3} B. Liang\altaffilmark{1} Y.B. Liu\altaffilmark{1, 2, 3} X.L. Zhang\altaffilmark{1} and K.F. Ji\altaffilmark{1}}
\email{wangfeng@acm.org}


\altaffiltext{1}{Computer Technology Application Key Lab of Yunnan Province, Kunming University of Science and Technology, Chenggong, Kunming, China, 650500}
\altaffiltext{2}{Yunnan Observatories, Chinese Academy of Sciences, Kunming, China, 650011.}
\altaffiltext{3}{University of Chinese Academy of Sciences, Beijing, China,100049.}
\altaffiltext{4}{Chinese National Observatory, Chinese Academy of Sciences, Beijing, China, 100011.}


\begin{abstract}
The Chinese Spectral RadioHeliograph (CSRH) is a synthetic aperture radio interferometer built in Inner Mongolia, China. As a solar-dedicated interferometric array, CSRH is capable of producing high quality radio images at frequency range from 400 MHz to 15 GHz with high temporal, spatial, and spectral resolution.To implement high cadence imaging at wide-band and obtain more than 2 order higher multiple frequencies, the implementation of the data processing system for CSRH is a great challenge. It is urgent to build a pipeline for processing massive data of CSRH generated every day. In this paper, we develop a high performance distributed data processing pipeline (DDPP) built on the OpenCluster infrastructure for processing CSRH observational data including data storage, archiving, preprocessing, image reconstruction, deconvolution, and real-time monitoring. We comprehensively elaborate the system architecture of the pipeline and the implementation of each subsystem. The DDPP is automatic, robust, scalable and manageable. The processing performance under multi computers parallel and GPU hybrid system meets the requirements of CSRH data processing. The study presents an valuable reference for other radio telescopes especially aperture synthesis telescopes, and also gives an valuable contribution to the current and/or future data intensive astronomical observations.
\end{abstract}



\section{Introduction}
\label{sec:intr}

The solar activities such as coronal mass ejections (CMEs), flares, and solar energetic particles (SEPs) have critical influence on space weather because of sudden energy release, particle acceleration, and/or transportation processes of the solar magnetic field. The observations of radio bursts are important diagnosing approach to reveal the related parameters, such as the magnetic field, electron density, plasma temperature and so on, to the various solar activities.

The Chinese Spectral RadioHeliograph (CSRH) is a synthetic aperture radio interferometer which is capable of observing radio bursts and producing high quality radio images at frequency range from 400 MHz to 15 GHz with high temporal, spatial, and spectral resolution. The goal of CSRH is to further understand the coronal dynamics.

\begin{figure}[!htb]
\centering
\includegraphics[width=0.4\textwidth]{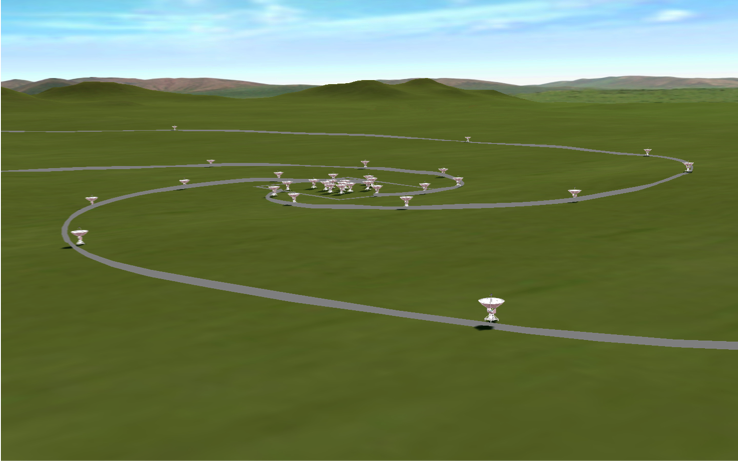}
\caption{The antenna distribution diagram of CSRH. }\label{AntennaDistribution}
\end{figure}

CSRH consists of total 100 radio antennas spirally distributed at Mingantu town, Inner Mongolia of China (see Fig.~\ref{AntennaDistribution}).
The RF signal of CSRH in 0.4 - 15 GHz is divided into 0.4 - 2 GHz (subarray CSRH-I), and 2 - 15 GHz (subarray CSRH-II) bands.
CSRH-I in 400 MHz - 2 GHz with 40 antennas of 4.5 meters, and CSRH-II in frequency range of 2 - 15 GHz with 60 antennas of 2 meters have been successfully installed and the first light image was obtained on Feb, 2013. The final specifications of CSRH, as driven by scientific goal, are shown in Table 1~\citep{YAN204,YAN2009,YAN2010,YAN2011}.

\begin{table}[!htb]
\small
\centering
\caption{The specification of CSRH\label{tblSpecification}}
\begin{tabular}{ll}
\hline  \hline
Frequency range & 0.4-15 Ghz ($\lambda$: $\sim$75 - 2 cm) \\
Frequency resolution & 64 channels (I: 0.4 - 2 GHz) \\	
& 528 channels (II: 2 - 15 GHz) \\
Antennas & I: 40 \\
& II: 60 \\
Baselines & I: 780 \\
&  II: 1770 \\
Correlation capacity & I: 780$\times$16 = 12480 \\
& II: 1770$\times$16=28320\\
Spatial resolution & $\sim$1.3" - 50" \\
Temporal resolution & I: 25 ms\\
& II: 206 ms  \\
Dynamic range & 25 dB (snapshot) \\
Polarizations & Dual circular L, R \\
Lmax & about 3 km \\
Field of view & 0.6$^{\circ}$ - 7$^{\circ}$ \\
\hline
\end{tabular}
\end{table}

The construction of  data processing system (DPS) for CSRH is a big challenge. Ideally, the DPS should deal with all the data produced by CSRH in real-time including data acquisition, data storage, data processing and data publication. However,
as a radio synthesis aperture  telescope with high temporal resolution, high spectral resolution and high spatial resolution, CSRH would produce massive observational data every day which are hard to be promptly processed. In every 3 ms, the digital receivers of CSRH-I and CSRH-II would generate a data frame which includes the auto-correlation and cross correlation data of 16 channels, and output to the specified computer respectively.
From point of view of data amount, CSRH-I would output about 31 MB data per second and about 1.05 TB in an observational day of 10 hours. The amount of data produced by CSRH-II would be approximate 2.146 TB in a day. In a month, the size of all observational data would be about 100 TB. Meanwhile, to monitor the operation of the full telescope system, it is necessary to generate images as quickly as possible so as to determine the status of the telescopes.

It is necessary and urgent to set up an automatic data processing system (i.e., pipeline) that is one of the most significant issues in the construction of CSRH. Although the amount of the observational data is far below the amount of Square Kilometre Array (SKA)~\citep{Beck2004,Dewdney2009}, CSRH is a solar radio interferometry that would produce the largest amount of data in the present. Therefore,  the high performance data processing system of CSRH is an valuable reference for other radio telescopes especially the aperture synthesis telescopes.

In this study, we present a distributed data processing pipeline (DDPP) built on our own-designed distributed computing infrastructure named OpenCluster. After the brief introduction of CSRH, we concentrate our study on the design of high performance data processing pipeline. The rest of this paper is organized as follows. Section 2 discusses the system architecture of CSRH. The requirements of CSRH data processing is discussed in Section 3. We introduce the OpenCluster in Section 4. The key techniques of CSRH pipeline are presented in Section 5. The performance of the DDPP is listed in Section 6. Finally, discussions and a short summary are provided in Section 7 and 8 respectively.


\section{CSRH Architecture}\label{sec:architecture}

\subsection {The Architecture of CSRH}
\begin{figure*}[htbp]
\centering
\includegraphics[width=0.8\textwidth]{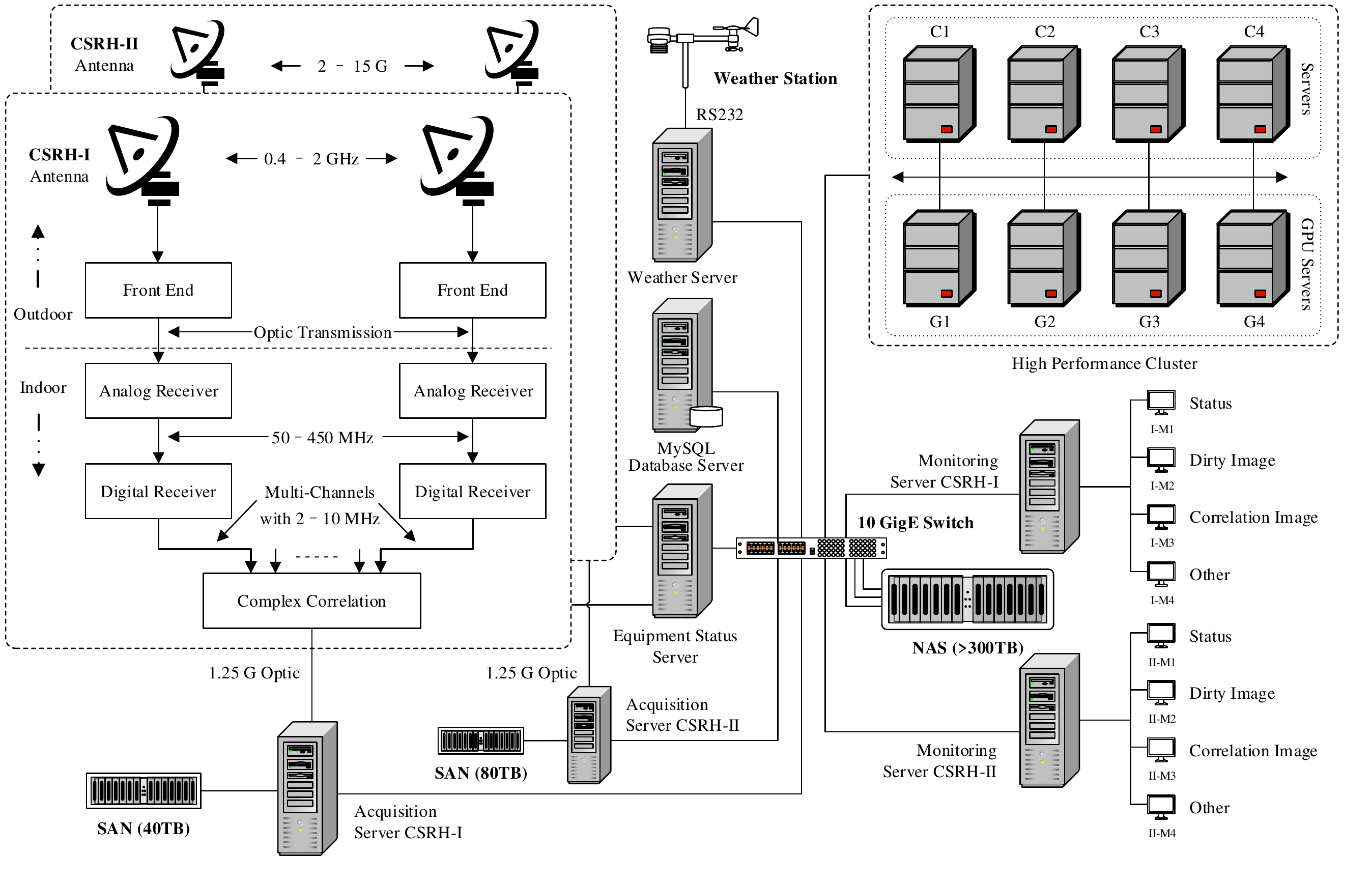}
\caption{The diagram of CSRH architecture. }\label{CSRH-architecture}
\end{figure*}

Fig.~\ref{CSRH-architecture} shows an architecture diagram of CSRH. The outdoors equipments consist of antennas, wide-band feeds, low noise amplifiers, optic transmitters, optic fibers, control units, as well as power supplies, and so on. The RF signal bands are transmitted to the indoor unit by optical fibers. The array configuration is chosen as a self-similar spiral geometrie~\citep{WangWei2011}.

The indoor equipments include optic receivers, analogous receivers, A/D converters, digital correlation receivers, local oscillators, monitors, computers, and so on (see Fig.~\ref{CSRH-architecture}). The signal of each band will be processed digitally. A 1 Gsps analog digital converter is used. The sampled signals then go through a digital spectral analyzer of 16 channels with 25 MHz being bandwidth of each channel for CSRH. Then signals are correlated at a frequency point within each channel. This procedure is repeated to cover the whole frequency so as to cover the whole channels. The signal from each 25 MHz will be correlated with signals from other antennas. The delay compensation bank is estimated to be in the range of 10 microseconds with step of 1 ns.

\subsection {The Architecture of Data Processing Environment}
A computer cluster environment is built for CSRH data processing (see Fig.~\ref{CSRH-architecture}). The system can efficiently realize the load balance, high compatibility and good extensibility. With the increase of computation load, the number of the server can be seamlessly expanded to obtain more computing power.

So far, the computer cluster consists of eight servers. Each server has 2-way Intel Xeon E5-2650 v2 CPUs, 2.6 GHz, 16 cores, 32 GB memory and 1 TB hard disk. Considering the processing requirements of massive data, the computer cluster is divided into two sub clusters. One sub-cluster (i.e., Cluster-C including C1 - C4) is mainly used for file processing. Another sub-cluster (i.e., Cluster-G including G1 - G4) is mainly for high performance imaging and all servers in Cluster-G are installed a NVIDIA Tesla C2050 Graphics Process Unit (GPU) card (http://www.nvidia.com) respectively.

A Network Attached Storage (NAS) system is deployed for CSRH-I/II observational data archive. The available space size of the NAS is about 300 TB and would be expanded to 1 PB in the near future.

All servers are connected to a 10 Gb Ethernet Switch using 10 Gb links. The NAS system is also connected to the 10 Gb Ethernet switch but using multiple 10 Gb aggregated links to guarantee the communication performance.

In addition, four types of dedicated servers are deployed.

1) Acquisition server is used to receive observational data from the digital receiver in every 3 ms and stores the data to the storage system. Meanwhile, the acquisition server would repeatedly forward the observational data to monitoring servers for real-time monitoring. The data forward frequency can be adjusted according to the monitoring requirements.

Two acquisition servers are deployed for CSRH-I and CSRH-II respectively. Each acquisition server has a stand-alone storage area network (SAN) device to temporarily store observational data. The available storage space of the SAN is considered to occupy the observational data for a month.

2) Database server is used to store parameter values. A MySQL database is installed on the server to store instrument parameters, telescope status and weather data.

3) Weather server is a gateway that acquires the weather data from a Vantage Pro weather station (http://www.davisnet.com/). It would acquire data in every 1 minute and save the weather information to the database server. Vantage Pro2 weather station currently installed measures barometric pressure, temperature, humidity, rainfall, wind speed and direction, UV/solar and so on.

4) Monitoring server is used to monitor running states via visualization method. A display adapter with four DVI output ports is installed on the server so as to connect four LCD monitors simultaneously. Two monitoring servers are in charge of status display for CSRH-I and CSRH-II respectively.

\section{Requirements Analysis}

\begin{figure}[htbp]
\centering
\includegraphics[width=0.49\textwidth]{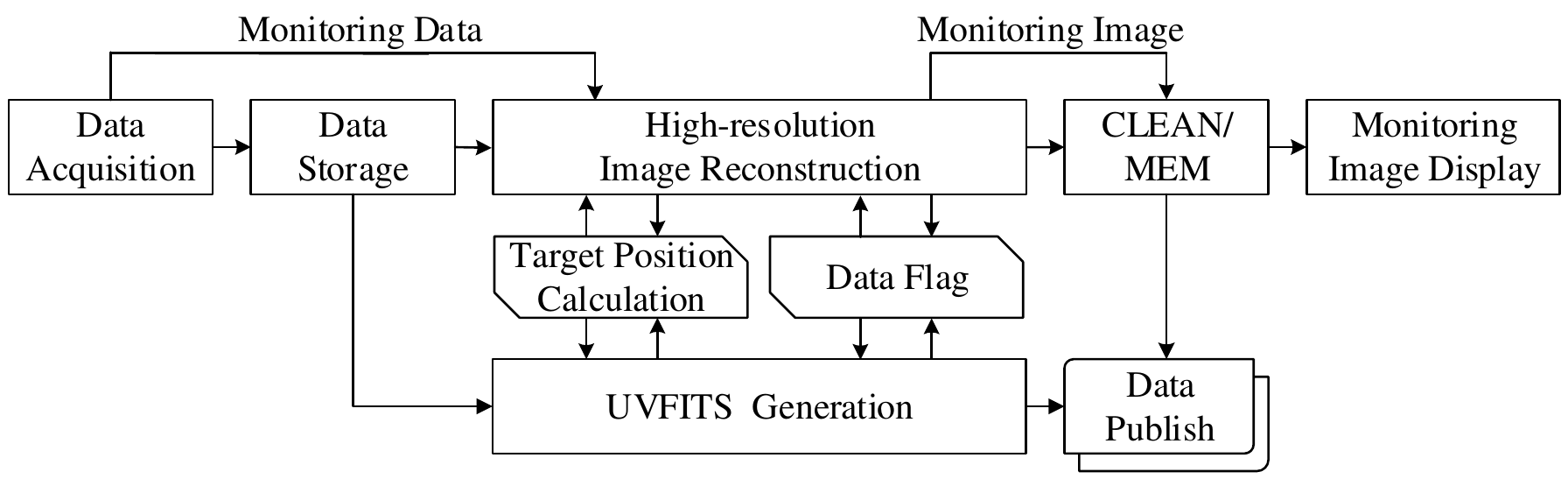}
\caption{The flowchart of CSRH data processing. }\label{CSRH-PreProcessing}
\end{figure}

The data processing of the synthetic aperture radio interferometry has been deeply described by previous literatures~\citep{Thompson2008,mcmullin2007casa}.  Fig.~\ref{CSRH-PreProcessing} shows the flowchart of CSRH data processing.

Obviously, all the procedures in Fig.~\ref{CSRH-PreProcessing} should be implemented in CSRH pipeline.  Meanwhile, referring to the related studies \citep{Jenness2014,Freudling2013,Shamir2008,Hummel2006} on pipeline design of modern telescopes, the requirements listed as follows are critically considered and designed because of the specific features of CSRH.

1) Data storage and data distribution. The storage of observational data is the premise of the data processing. Current CSRH observational data are saved to the SAN system firstly. However, due to the performance limitation of the SAN, it is hard to synchronously read observational data from the SAN while writing data. This means we cannot read data from the SAN while in CSRH observation. The observational data have to be read and processed in batch after the observation every day.

2) Data archive format.  The data format of the observational data for archiving is a worthwhile problem in CSRH data storage. Due to the amount of the observational data, the different data storage format would significantly affect the hardware configuration and the available space of the storage system. The available space of storage system is closely related to the archive format. In addition, the high performance index is another important issue while designing high performance data processing system of CSRH, or the data retrieval would be a bottle-neck of CSRH data processing.

3) High performance imaging.  CSRH is capable of observing 64 channels in CSRH-I and 528 channels in CSRH-II. The high performance imaging is an urgent demand for data processing, publication and monitoring. However, the deconvolution manipulation for dirty images is a very time-consuming calculation procedure. It is necessary to develop high performance imaging technique so as to improve the scientific output of CSRH.

4) Customizable workflow.  With the change of scientific research goal, the data processing flow would also be changed. The DDPP should support customizable workflow in data processing so as to dynamically adjust the data processing flow and satisfy the requirements of astronomical scientists.

5) Data reprocessing. Data reprocessing is the critical requirement of CSRH.
Due to the change of requirements, the observational data often need to be processed in different ways by scientific researches. For example, to improve the spatial resolution, the scientists need to integral multi-frame data in any given period of observational time.

\section{OpenCluster - Distributed Computing Infrastructure}
To design and develop the DDPP, we first develop a novel distributed computing infrastructure named OpenCluster, which is a wholly own-designed software for quickly designing scientific data processing pipeline.

Referring to the design and operation principles of stream computing~\citep{neumeyer2010s4,buck2004brook}, the OpenCluster simplifies the technical implementation of stream computing, and further adds greater design capability and many additional features for astronomical data processing.

\begin{figure}[htbp]
\centering
\includegraphics[width=0.4\textwidth]{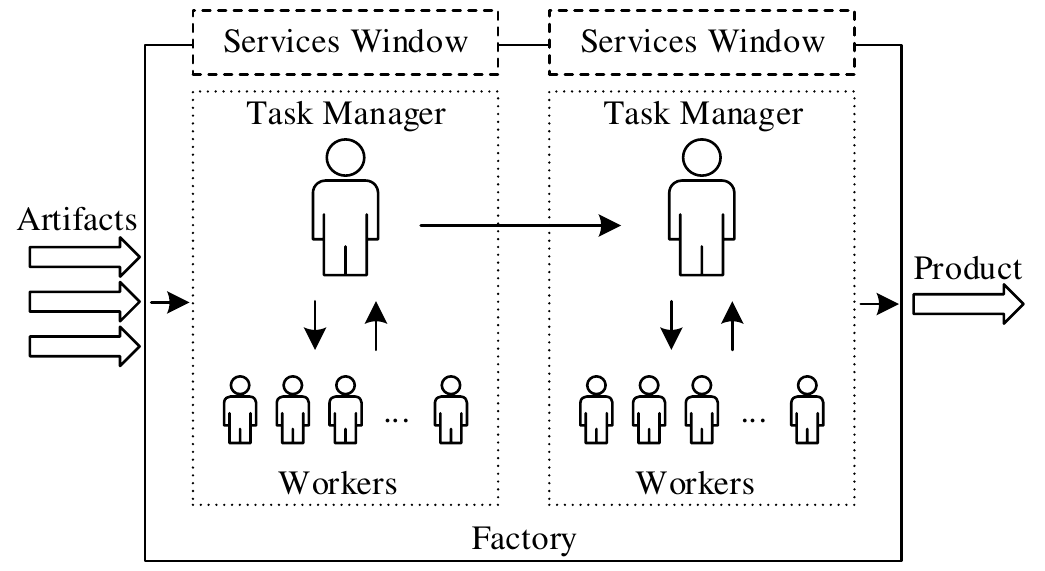}
\caption{The conceptual diagram of OpenCluster. }\label{Fig_cfactory}
\end{figure}

Fig.~\ref{Fig_cfactory} shows a conceptual diagram of the OpenCluster. The OpenCluster regards a data processing pipeline as a data  processing factory. The factory is in charge of data processing and undertakes all data processing tasks. In the factory, there are many task managers who manage a group of workers. The task managers obtain the tasks from the factory or other task managers,  then schedule their subordinate workers to run the tasks, and finally collect the task results from the workers. In addition, there are several external service windows opened for all workers and task managers in the factory to provide specified services.

\begin{figure*}[htbp]
\centering
\includegraphics[width=1.0\textwidth]{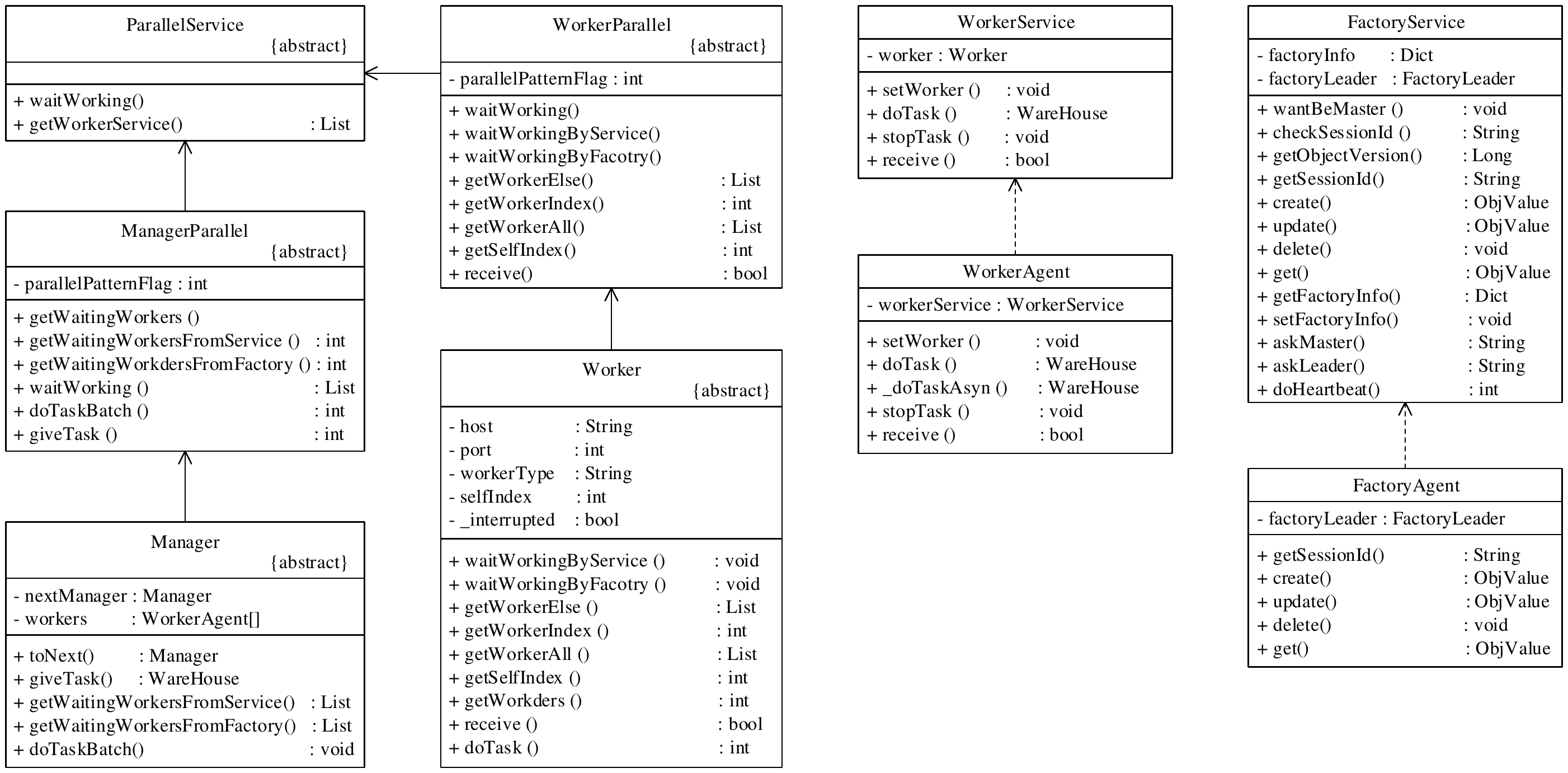}
\caption{The class diagram of the OpenCluster. }\label{Fig-ClassOpenCluster}
\end{figure*}

It is easy to design astronomical data processing pipeline using the OpenCluster infrastructure. The OpenCluster is written with Python language which is widely used in astronomy. Many mature packages such as PyFits~\citep{barrett1999pyfits} and Pyro4 (https://pythonhosted.org/Pyro4/) are used in the software development. The OpenCluster is a pure Python application which can be installed on any operating system. The class diagram of the OpenCluster is shown in Fig.~\ref{Fig-ClassOpenCluster}.

Current OpenCluster edition has encapsulated all complex concepts of distributed computing, e.g., task schedule, task distribution, message passing and so on. The software developers do not need to care the implementation of the OpenCluster and distributed computing knowledge, then only need to implement the codes of inherited classes such as different workers and managers.

The ZeroMQ~\citep{hintjens2013zeromq} is used for data and command communication among all components including factory, global services, task managers, task workers. With the support of the ZeroMQ message queues technique, the OpenCluster is robust to transfer data, messages and commands communication, and finally guarantee the robust and availability of the full system.

To implement workflow mechanism of CSRH pipeline, data-driven mode is used in OpenCluster. All observational data  processed in the pipeline would be add a tag to identify the properties of the data. The form of data tag is $<$data source : process mode : publish mode$>$. $<$data source$>$ represents the type of the data. $<$process mode$>$ represents processing approaches for observational data.

\section{The Implementation Of The DDPP}

According to the data processing flow of CSRH (see Fig.~\ref{CSRH-PreProcessing}), we create six task modules and four services that inherited from the base classes of the OpenCluster.

\begin{figure}[htbp]
\centering
\includegraphics[width=0.5\textwidth]{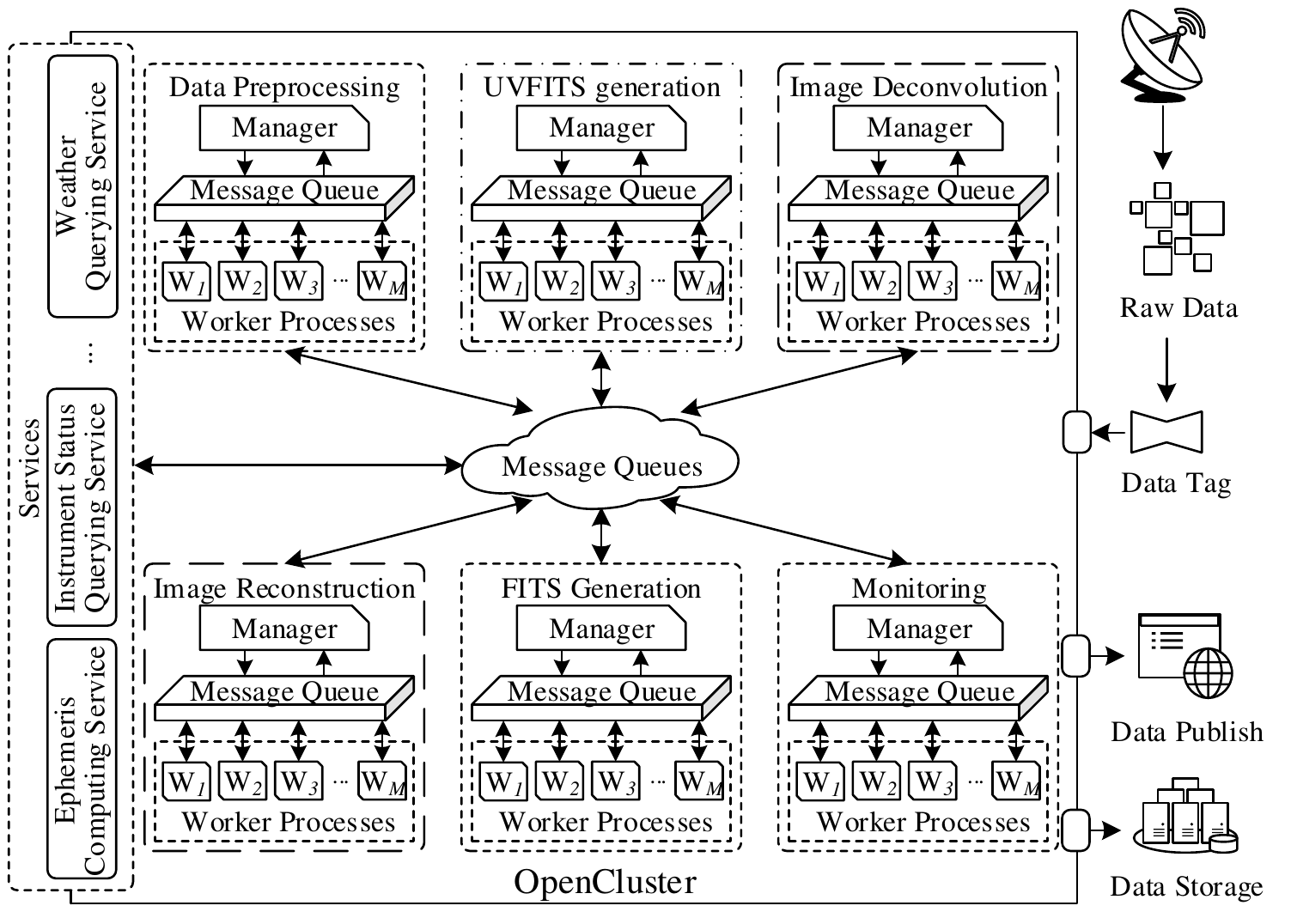}
\caption{The structure diagram of the DDPP based on the OpenCluster infrastructure. }\label{Fig_factory}
\end{figure}

Fig.~\ref{Fig_factory} shows a structure diagram of the DDPP. The six task modules include the fundamental functions of data preprocessing, image reconstruction, image deconvolution, raw observational data UVFITS file generation, FITS file generation for final data products, and observation monitoring. Meanwhile, three global services are designed to provide the services of ephemeris computing, instrumental status query and weather information query.

\subsection{Workflow Tag Design}
The OpenCluster provides some level of support for workflow technology. According to the data tag, OpenCluster would schedule the proper managers to process the data and finally obtain the scientific data production. The DDPP is a standalone computing platform which supports processing CSRH-I and CSRH-II observational data simultaneously. Therefore, the design of workflow tag is important for the DDPP. Meanwhile, these tags can be expanded according to the scientific requirements in the future.

Table~\ref{tblTag} lists all possible tags including 6 data sources, 5 process modes and 4 data publication modes in the present. It is easy to understand the definition of the tag.
For example, the data tag of $<$1:5:3$>$ means that the input data are real-time observational data from CSRH-I, these data should be firstly pre-processed, and then be generated to the dirty images and further be performed deconvolution. Finally, these images would be published with FITS file format.

\begin{table*}[htbp]
\small
\centering
\tabcolsep 3pt
\renewcommand{\arraystretch}{1.3}
\vspace*{2mm}
\caption{The form of data tag\label{tblTag}}
\begin{tabular}{l|l|l}
\hline \hline
Data Source & Process Mode & Publish Mode \\
\hline
1. CSRH-I real-time data          & 1. Preprocessing           & 1.  UVFITS File\\
2. CSRH-II real-time data     & 2. Dirty image       & 2. FIT-IDI File\\
3. CSRH-I batch data      & 3.  Deconvolution         & 3. FITS File\\
4. CSRH-II batch data    & 4. 1+2                                  & 4. PNG File\\
5. CSRH-I batch integral data  & 5. 1+2+3                              &\\
6. CSRH-II batch integral data&                                             & \\
\hline
\end{tabular}
\end{table*}

\subsection{Data acquisition and distribution}
To acquire the massive data produced by CSRH, the acquisition server receives the observational data encapsulated as a frame from digital receivers through a 1.25 Gb optical fibre in every 3 ms. A frame includes the observational data of 16 channels with 1 polarization. Therefore,  the acquisition server has to receive multi consecutive frames to acquire all channels and all polarizations. For CSRH-I, the acquisition server should receive 8 consecutive frames to generate a full frame (64 channels and 2 polarizations) and would take 25 ms ($3\times8+1$ ms data read out ). The data of CSRH-II is close to CSRH-I. The major difference is that CSRH-II has 528 channels. Hence, the size of a frame in CSRH-II is 204,800 bytes and total 33 frames are treated as a full frame. The acquisition period is 206.25 ms.

\begin{figure}[htbp]
\centering
\includegraphics[width=0.49\textwidth]{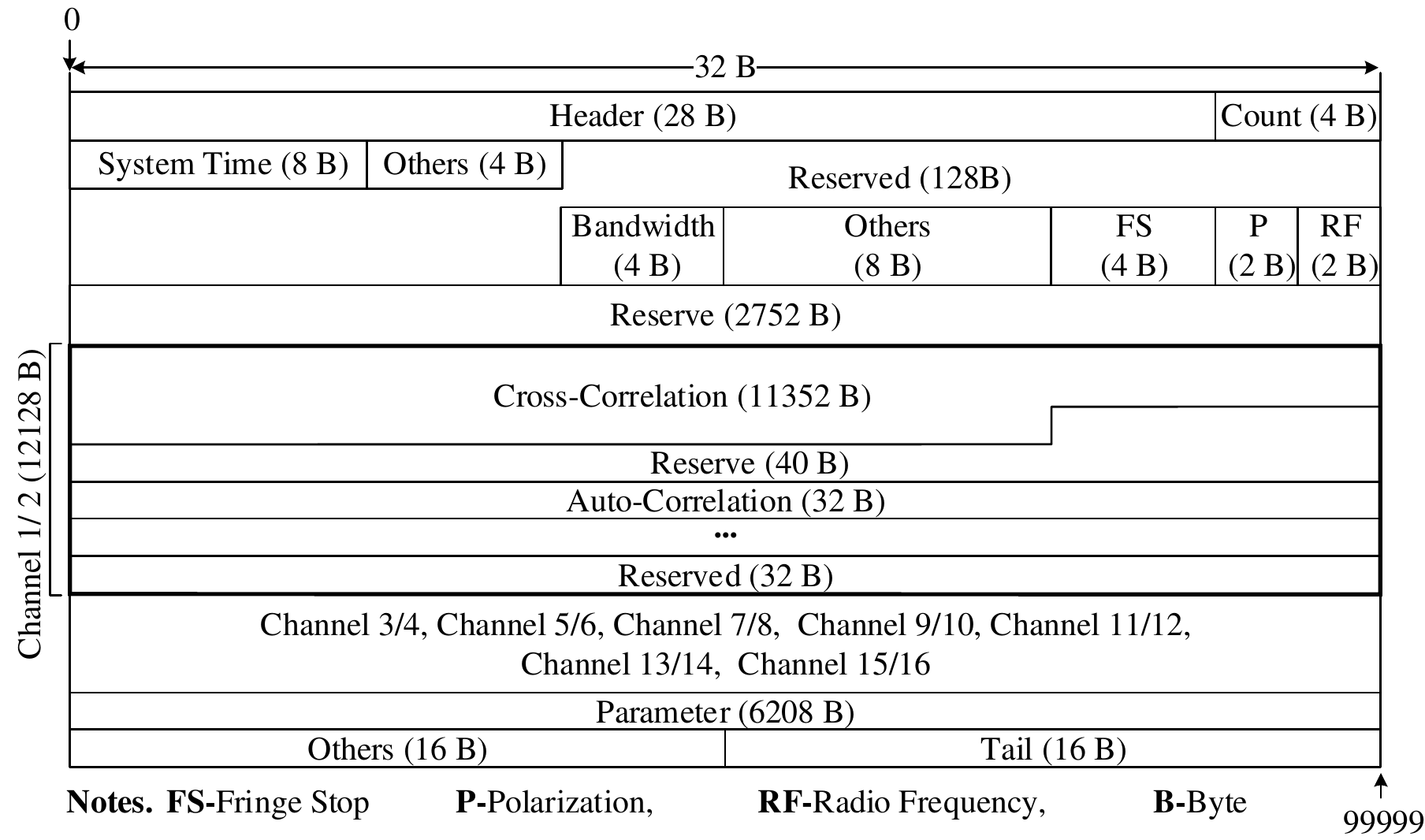}
\caption{The diagram of CSRH Data Format. }\label{Fig-CSRH-Frame}
\end{figure}

The raw data format definition of CSRH-I is shown in Fig.~\ref{Fig-CSRH-Frame}. All observational parameters such as polarization, band and channel are stored and can be easily readout according to the pre-defined byte stream offset. The format of CSRH-II is close to the format of CSRH-I.

To monitor the instrumental status, we have to periodically extract parts of observational data and send to the monitoring system via TCP protocol. The monitoring server acts as the TCP server.

Considering the requirements of observation monitoring, 5 seconds in CSRH-I and 15 seconds in CSRH-II are setup as the sampling period. In every sampling period, the data acquisition server will send 16 frames (CSRH-I) and 66 frames (CSRH-II) to the monitoring server. And the monitoring server would separate a full frame from these frames for subsequent data processing.

\subsubsection{Observational Data storage and file format}
After comparing the advantages and disadvantages of each file format, we determine to archive the observational data with CSRH raw data format. Actually, most modern telescopes use specified data format to store observational data. For example, The ALMA and JVLA projects store the data with a common archival science data model (ASDM) format~\citep{glendenning2008alma}, and have jointly developed the software to fill this data into CASA~\citep{mcmullin2007casa}. In the ASDM format, the bulk of the data is contained in large binary data format (BDF) tables, with the meta-data and ancillary information in XML tables. Meanwhile, due to the wide applications of CASA software, Measurement Sets (MS) format is also a common file format in radio astronomy.

The storing with ASDM or MS format would bring more convenient for further data sharing and utilizing, but it would occupy many additional storage spaces. For a frame acquired in every 3 ms of CSRH-I, the size is 100,000 bytes. Table~\ref{tblFormat} lists the size with different file format respectively. Obviously, due to the using of XML and meta data definition, the size of ASDM and UVFITS files would be significantly increased and further increase the expenditure of the storage system. For CSRH and its massive observational data, it is a huge stress burden because of limited funds.

\begin{table}[htbp]
\small
\centering
\tabcolsep 3pt
\renewcommand{\arraystretch}{1.3}
\vspace*{2mm}
\caption{The file size of each common file format in radio astronomy\label{tblFormat}}
\begin{tabular}{ll}
\hline \hline
File Format & File Size (Bytes) \\
\hline
RawData & 100,000 \\
UVFITS & 200,000 \\
FITS-IDI & 368,000 \\
Measurement Set & 2,200,100 \\
ASDM & 324,000 \\
\hline
\end{tabular}
\end{table}

\subsubsection{Data Archive and High Performance Index}

{\em 1. Observational Data Archive}

\begin{figure}[htbp]
\centering
\includegraphics[width=0.45\textwidth]{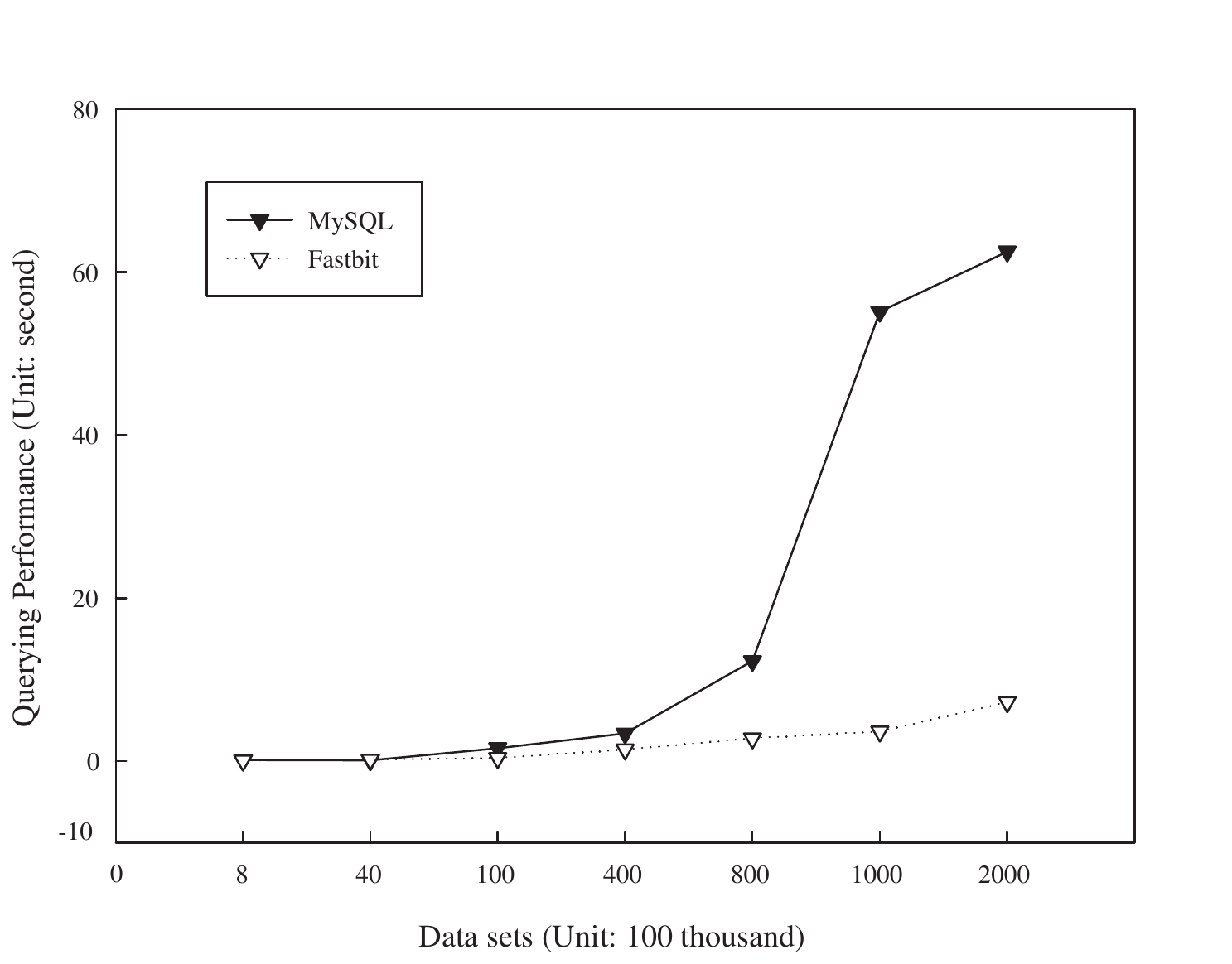}
\caption{The diagram of comparison between Fastbit and MySQL database. }\label{Fig_Fastbit}
\end{figure}

The data index technique is critically significant for CSRH to quickly retrieve observational data from massive data of CSRH. So far, all observational data are saved into the storage system in file form. About 12 million observational data frames in a day, and about  billion frames in a month would be output by CSRH-I. To quickly retrieve a specified frame and locate the corresponding file, the relational database technology has been widely used to manage the index information of the observational data. However, based on our preliminary experiment,
we realize that it is hard to archive so many observational data files and meet the performance requirements of the subsequent data processing (see Fig.~\ref{Fig_Fastbit}). The query performance of MySQL under more than 0.1 billions records would take more than 60 seconds to fetch a record. Obviously, this result would critically limit the processing performance of the DDPP.

We create indexes for observational data  by using Fastbit~\citep{Wu2005,Wu2009, LYB2014} technique.
Fastbit is very well suited for managing massive data because of its bit index technique. As an open-source data processing library following the spirit of NoSQL movement, Fastbit offers a set of searching functions supported by compressed bitmap indexes~\citep{Wu2001}. It treats user data in the column-oriented manner.

The Fastbit indexes for each observational data file are automatically created while transferring the observational data from SAN to NAS.
The main index fields include file name and its location, observational date and time, polarization, band and frame byte offset. Frame byte offset means the byte offsets of a specified frame from the beginning of the observational file. In data processing, it is easy to retrieve the information of file name, location and frame byte offset with the query parameters of observational date, time, polarization and band. The subsequent program can open the observational file with the retrieved file name, skip the bytes defined by the frame byte offset, and directly read the observational data needed.

{\em 2.  Parameter Data Archive}

Besides the archiving of observational data, it is necessary to separately record all parameter data, such as weather conditions, instruments status (i.e., antenna, receivers, and so on), instruments parameters (the position of each antenna, the length of each optical fibre, and so on) in a timeframe. To guarantee the correspondence of CSRH observational data, these data have to be permanently stored and can be retrieved according to the observational date and time.

\begin{figure}[htbp]
\centering
\includegraphics[width=0.4\textwidth]{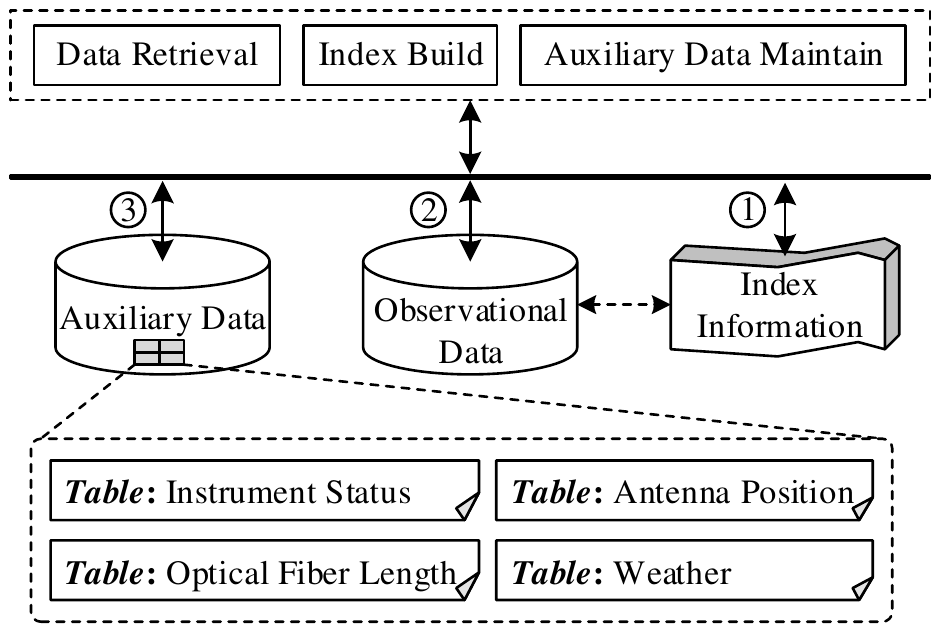}
\caption{CSRH data archive, index and the deployment of related databases. }\label{CSRH-Database}
\end{figure}

All parameter data are stored in a Mysql database. Four tables such as instrumental status, optical fibre length, antenna position and weather are created (see Fig.~\ref{CSRH-Database}).
1) Instrumental status table records the status of the telescope, especially the availability of each antenna that can be used to flag the observational data in data processing. 2) Optical fibre length table is used to record the length of each optical fibre between the outdoor and indoor devices which would be used in computing RF signal transfer delay. Although the length is rarely varied, it is necessary to record for high precision computing. 3) Antenna position table stores all the locations and the altitude of all antennas. The center position of CSRH is  (0., 0.). And the deviation values from the center position of the each antenna are stored in the table respectively. 4) Weather conditions table records the weather information.

There are total 3 steps to retrieve the observational data and the related parameters (see Fig.~\ref{CSRH-Database}). Step 1, the DDPP reads the observational date and time from the raw data. Step 2, the DDPP retrieves the observational data by using Fastbit index and then locate the file directory. Step 3, the DDPP retrieve the parameters from four tables according to the date and the time respectively.

\subsection{Tasks Modules}
\subsubsection{Data Preprocessing}
Data preprocessing is a significant part in CSRH data processing. The goal of data preprocessing is to correct, flag, compensate and calibrate the observational data.
Meanwhile, it is possible to overlap the data among several continuous frames in the data preprocessing to generate an integral image with higher resolution.

{\em 1. Delay compensation and Fringe Stop}

The delay compensation and the fringe stop must be conducted in CSRH data processing. The digital receiver of CSRH would encapsulate
 the values of transmission delay into the frame and output to the acquisition server. Therefore, to correct the transmission delay and conduct fringe stop, the follow steps are implemented in data preprocessing.

Step 1: Obtain the correlated visibility data from the observational data, including the real and imaginary parts, define as $Ae^{j\phi}$, where A is the amplitude, $\phi$ is the phase.

Step 2: Obtain the delay parameter ($dt_{raw}$) from the observational data encapsulated by the digital receiver.

Step 3: Subtract delay parameter from the delay skew ($dt_{trans}$) caused by different length of the optical fibre. These delay skews are recored in the database.

Step 4: Compute the delay between antenna i and j: $dt_{ij}=dt_j-dt_i$, where $dt=dt_{raw}-dt_{trans}$. The corresponding correlation value is $A_{ij}e^{j\phi}$.

Step 5: Obtain the observation frequency ($F_{rf}$) and intermediate frequency ($F_{if}$) of each channel from the observational data.

Step 6: Compute fringe stopping. The phase, which could be compensated on the complex correlation value $A_{ij}e^{j\phi}$ between antenna i, j, can be defined as:
$\Phi_{fs\_ij} = 2\pi(F_{rf} \times{}dt_{ij} - F_{if} \times (dt_j - dt_i))
$, then, the complex correlation value can be computed by subtracting initial phase from the phase of fringe stopping: $A_{ij}e^{j\phi}=A_{ij}e^{j (\phi - \phi_{fs\_ij})}$.

{\em2. Satellite calibration}

CSRH observes satellite to calibrate the phase of each channel. The calibration steps are listed as follows.

Step 1:  Obtain the correlated visibility function from the observational data, including real and imaginary parts, define as $A_{sun}e^{j\phi_{sun}}$, where $A_{sun}$ is the amplitude, $\phi_{sun}$ is the phase.

Step 2: Obtain the correlated visibility data from the observational data, including real and imaginary parts, define as $A_{satellite}e^{j\phi_{satellite}}$, where $A_{satellite}$ is the amplitude, $\phi_{satellite}$ is the phase.

Step 3: Obtain the result: $A_{sun}e^{j\phi_{sun}}=A_{sun}e^{j(\phi_{sun}-\phi_{satellite})}$.

{\em3. Data Flagging}

Radio Frequency Interference (RFI) is a disturbance that affects an electrical circuit due to either electromagnetic conduction or electromagnetic radiation emitted from an external source. Especially at low radio frequencies of CSRH, stray electromagnetic transmissions often interfere with the incoming radiation from a source, thus corrupting the data being recorded.
Therefore, flagging involves the identification and masking of RFI affected data points, and is an inevitable step in standard data analysis.

The DDPP supports two main data flagging approaches. One is using hardware information which acquired from the instrumental status to directly flag bad data. Another is using automatic RFI Identification and flagging algorithm that has been integrated into CASA software~\citep{urvashi2003automatic}.

\subsubsection{Monitoring}
Two servers are specifically deployed for monitoring CSRH-I/II observation in real-time respectively. After the task of data preprocessing, the  results would be transferred to monitoring task module. So far, to monitor the status of each antenna and its corresponding parameters, we refer to the monitoring functions of other interferometers, and design two types of diagrams for real-time monitoring. One is power and phase scatter diagram. Another is the histogram diagram of auto correlation.

\begin{figure}[htbp]
\centering
\includegraphics[width=0.22\textwidth]{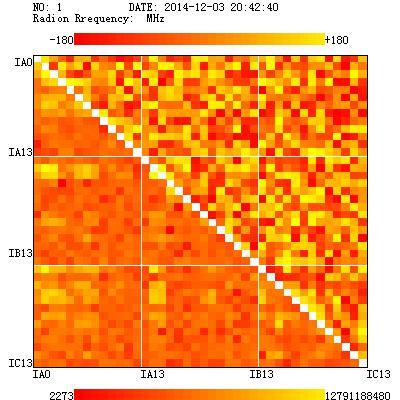}
\includegraphics[width=0.22\textwidth]{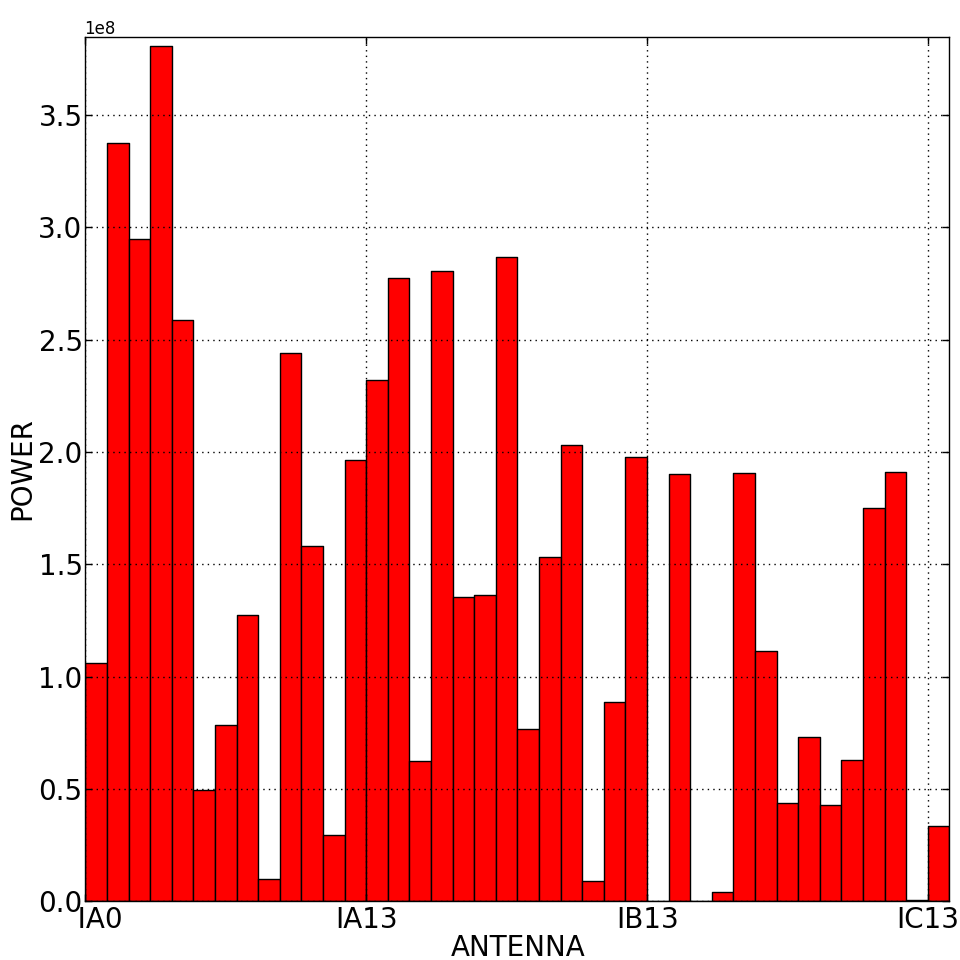}
\caption{The monitoring diagrams of CSRH. }\label{figMonioring}
\end{figure}

The power and phase scatter diagram (see Fig.~\ref{figMonioring} (left)) is for monitoring the baseline and its visibility data. The x and y axis is the number of each antenna. The power spectrum of each baseline is plotted in the bottom left corner of the diagram, and the phase is plotted in the up right corner. Obviously, if an error occurs during observational, the power spectrum of the corresponding baseline should be unusual. Therefore, a black line would be displayed in the power and phase scatter diagram.

The auto correlation diagram is a histogram diagram (see Fig.~\ref{figMonioring} (right)) which is used to monitor the auto correlation variations of each baseline. The x axis is the number of antenna and the y axis is the power spectrum of the autocorrelation.

\subsubsection{Distributed UVFITS/FITS-IDI File Generation}

It is necessary to generate UVFITS/FITS-IDI format files when CSRH needs to share observational data with other scientific group. The raw observational data cannot be directly processed by the other groups because the raw observational data have no corresponding information of observation. For example, the antenna status should be used to flag the observational data.

The DDPP should be capable of generating UVFITS and FITS-IDI formats. According to the official definition of UVFITS format~\citep{Wells1981,greisen2012aips}, the DDPP writes four binary tables (i.e., Primary HDU, AIPS FQ, AIPS AN andAIPS SU) to the UVFITS file. For FITS-IDI format file, the DDPP would write five binary tables such as ANTENNA, FREQUENCY, SOURCE, ARRAY\_GEOMETRY, UV\_DATA besides the primary HDU~\citep{greisen2011fits}.

\begin{figure}[htbp]
\centering
\includegraphics[width=0.4\textwidth]{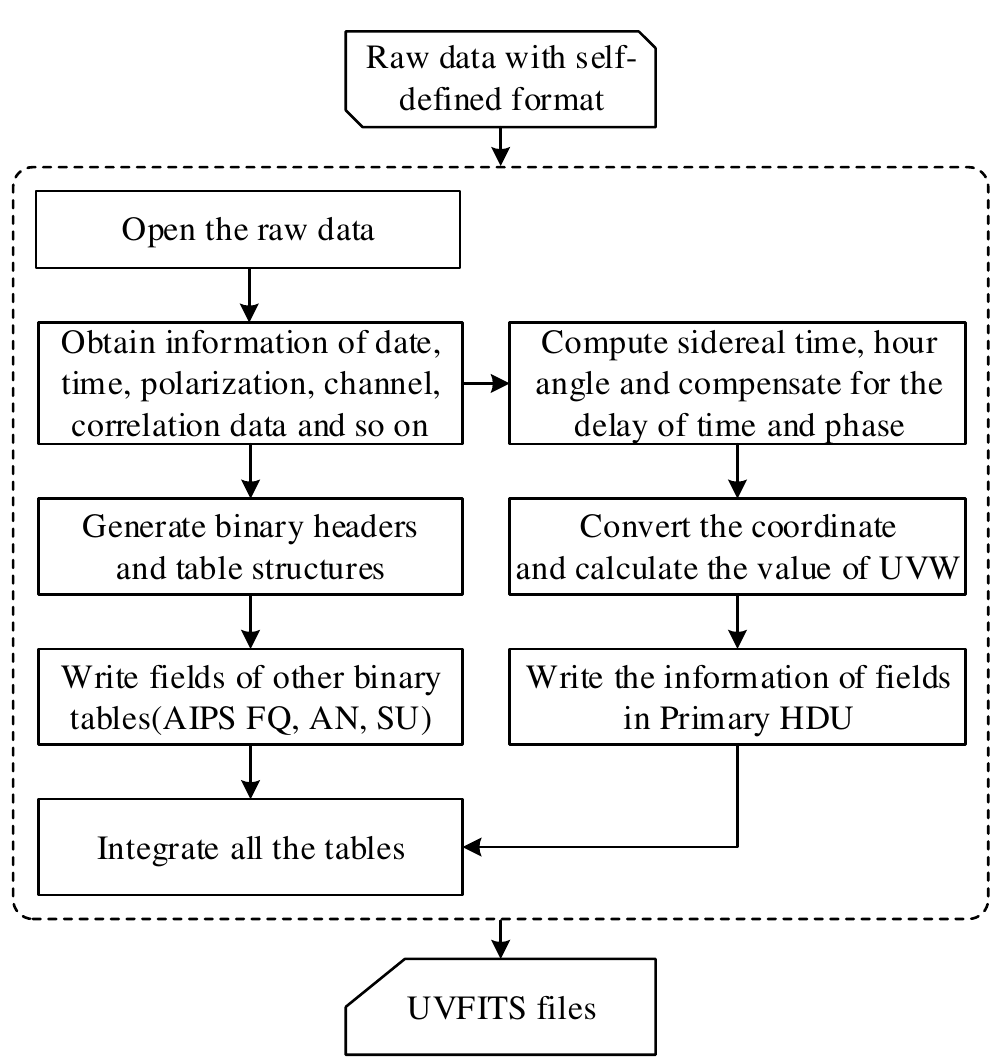}
\caption{Data flow char of UVFITS generation. }\label{CSRH-UVFITS}
\end{figure}

The implementation of UVFITS/FITS-IDI file generation module is very simple. However, to quickly generate a large amount of files, the DDPP used multi-process technique and can schedule several workers to generate files in the same time so as to improve the generation performance. The flow chart of UVFITS generation is shown in Fig.~\ref{CSRH-UVFITS}.

\subsubsection{High Performance Imaging And Deconvolution}
\label{sec:pipeline}

The H\"ogbom CLEAN algorithm~\citep{hogbom1974aperture} is used by CSRH data processing and other CLEAN algorithms are under developing. Due to the low performance of CLEAN algorithm, it is necessary to improve the performance of CLEAN algorithm.

The migration of CLEAN algorithm from CPU to GPU is feasible. Both the gridding and CLEAN kernels were parallelized by pixel (in the uv and image plane respectively). Meanwhile, once the maxima has been located in the dirty image, the convolution is embarrassingly parallel for both building the CLEANed image and subtracting from the dirty image.

Referring to an implementation of H\"ogbom Clean algorithm under GPU CUDA platform (http://nesanders.github.io/gICLEAN/index.html), we migrate the MIRIAD~\citep{sault1995retrospective} and its implementations from CPU to GPU environment.
We have implemented standard algorithms for both these tasks on the GPU, achieving speedups of $\sim$5 for gridding and $\sim$50 for CLEANing. The program is supported by PyCuda (http://mathema.tician.de/software/pycuda/) and Scikits.cuda (http://scikits.appspot.com/cuda).

\subsection{Global Services}
\subsubsection{High Precision Ephemeris Calculation}

High precision ephemeris calculation is the fundamental issue of CSRH processing system. Whether in  observation or in data processing, the position of the target (i.e., Sun or artificial satellite) is always an important parameter. Many processing procedures in CSRH data processing are seriously depended on the high precision target position.

Based on the evaluation of the baseline of CSRH, the largest baseline is about 3200 meters. It means the position accuracy should be superior to 1 macro arc-second while the observational frequency is 15 GHz~\citep{YAN2009}.

{\em1. JPL DE405 And Ephemeris Calculation.}

To obtain such accuracy level, the precise ephemeris must be considered. In the study, JPL DE405 planet ephemeris~\citep{Standish1998,Newhall1983,Charlot2005} is selected to provide high precision fundamental ephemeris. We select Naval Observatory Vector Astrometry Software (NOVAS)~\citep{kaplan2012novas} to construct a high precision ephemeris program for CSRH.

The computing procedures are as follows.

1)	Computing the X, Y and Z of the observation station in ECEF coordinate system based on the observational time (UTC), longitude (L), latitude (B) and the altitude (H) of the observation station.

2)	Calculate the position and the velocity of the observational target under International Celestial Reference System (ICRS) and the J2000.0 mean equatorial system of coordinates respectively. Meanwhile, the distance between the target and the observational station is also calculated.

3)	Calculate the position of the observational target at the observational time.

4)	Calculate the local apparent sidereal time of the topocentric coordinates. And further calculate the right ascension and the declination under the topocentric coordinate systems.

5)	Calculate the Greenwich sidereal time and further calculate the Local Apparent Sidereal Time (LAST) by using the geographic longitude.

{\em2. Automatic IERS Data Updating}

In ephemeris calculation, {\em TAI-UTC} and three earth orientation parameters (i.e.,  {\em x, y} and {\em UT1-UTC}) are necessary for high precision position calculation. To guarantee the precision of the ephemeris calculation, it is necessary to update these parameters from an official organization. We finally select the data from International Earth Rotation and Reference Systems Service (IERS - http://www.iers.org). The IERS was established in 1987 by the International Astronomical Union and the International Union of Geodesy and Geophysics. The IERS provides data on Earth orientation, on the International Celestial Reference System/Frame, on the International Terrestrial Reference System/Frame, and on geophysical fluids. It maintains also Conventions containing models, constants and standards.

The IERS publish four bulletins, i.e., Bulletin A contains rapid determinations for earth orientation parameters, Bulletin B contains monthly earth orientation parameters, Bulletin C contains announcements of the leap seconds in {\em UTC}, and Bulletin D contains announcements of the value of {\em DUT1}. To retrieve the parameters needed, we choose Bulletin A as the data source.

A linux daemon, that is named IERSSync,  is designed and run on the background of the server to automatically maintain the IERS data. The IRESSync is similar to a search engine crawler and would visit IERS website every day. After the retrieval of the web site pages, the IERSSync can analyze the contents of the HTML pages and try to locate the new IERS Bulletin A file URL. If new Bulletin A file published, the IERSSync will download the text file automatically, search the information from the text file and finally save the parameter data into the MySql database.

{\em3. High Performance Interpolation Computing}

The ephemeris calculation is a time-consuming task and the computing performance of ephemeris calculation is far below the expectation of real-time processing. Even on a high-end server with Intel Xeon 16 $\times$ 2.60 GHz cores and 32 GB memory, the calculation speed of one planet is about 20 ms. Obviously, it will cost large amount of time when processing single frame data of every 3 ms.

To guarantee the precision of the final position and obtain the maximum performance of computing, we finally use interpolation method. We use NOVAS to calculate a series of accurate positions (e.g., the $RA$ and the $DEC$ in UTC time {\em - 0:30, 0:00, 0:30, ..., 10:00}, and so on) of the observational target in a day, and interpolate the $RA$ and the $DEC$ to the given time. Obviously, the interpolation method should bring faster performance than the NOVAS.

We conduct several preliminary experiments to assess the availability and the performance upon two interpolation methods such as linear interpolation and 3-point parabolic interpolation. We regard that the final precision of interpolation would be greater than 0.001 arc second. The preliminary experiments results show that both interpolations methods can meet the requirements of the computing performance.  In the condition of the precision $> 0.001$ arc second, the linear interpolation method needs 49 accurate values (every 30 minutes) to compute the planet position by any given time in a day, and the parabolic interpolation method needs only 25 (every 1 hour) real values.

The DDPP use parabolic interpolation method to obtain high prevision position of the target. We statistic total time overhead of two interpolation methods and the related data initialization respectively, and finally choose parabolic interpolation method. Actually, the computing performances of two methods are very close in the high-end computer server. The main difference is the time overhead of data initialization. Therefore, the calculation of 25 real values by NOVAS only takes about 500 ms.

\subsubsection{Weather And Instrumental Status Query}
Due to the requirements of data flagging, the DDPP provides two global services to query weather information and instrumental status respectively. As mentioned in the previous section, we store these information in a MySQL database. Hence, the implementation of these two services is quite simply. According to the data and time, the service would compose a SQL statement and submit to the MySQL database. The query results would send back to the invokers.

In general, the weather condition would not affect the observation of CSRH. The DDPP only consider the two conditions that would affect the observational data such as strong wind and heavy rain.

\section{System Deployment And Application}

The DDPP has been deployed on each server for CSRH instrument testing and observation. All servers are installed CentOS 64 bits operation system. The version of Python is 2.9.7. All source codes of the DDPP are stored in a directory.

To test the availability of the DDPP, we focus on the two aspects. One is the correctness of the DDPP. Another is performance of data process. It is easy to verify the correctness of the DDPP because there are many mature and standard data processing software for synthetic aperture interferometer such as CASA and MIRIAD. We generate the UVFITS file of the observational data, import the file to CASA, and finally compare the results between the DDPP and the CASA.

Another significant issue of the DDPP is the computing performance. We carefully test the time overheads of each processing task that runs under one process and list the results in Table~\ref{tblTimeOverhead}.

 \begin{table}[!htb]
\small
 \centering
\tabcolsep 3pt
\renewcommand{\arraystretch}{1.3}
\vspace*{2mm}
\caption{The time overhead of data processing with one process\label{tblTimeOverhead}}
\begin{tabular}{l|l|c}
\hline \hline
Task & Sub Task & TO \\
\hline
\multirow{3}{2cm}{Data Preprocessing} & Analyzing one frame & 0.188\tabularnewline
\cline{2-3}
 & Delay compensation & 0.183\tabularnewline
\cline{2-3}
 & Computing UVW & 0.072\tabularnewline
\hline
\multirow{2}{2cm}{UVFITS Generation} & Creating tables & 0.251\tabularnewline
\cline{2-3}
 & Writing a UVFITS file & 0.026\tabularnewline
\hline
\multirow{2}{2cm}{Integral } &  One frame overlay & 0.031\tabularnewline
\cline{2-3}
 & Average value computing & 0.039\tabularnewline
\hline
\multirow{3}{2cm}{Gridding and Dirty map} &One 1024$\times$ 1024 image & 1.307 \tabularnewline
\cline{2-3}
 & One 512$\times$ 512 image & 1.005\tabularnewline
\cline{2-3}
 & One 256$\times$256 image & 0.860 \tabularnewline
\hline
\multirow{3}{2cm}{CLEAN with one iteration} &One 1024$\times$ 1024 image & 1.650\tabularnewline
\cline{2-3}
 & One 512$\times$ 512 image & 0.945 \tabularnewline
\cline{2-3}
 & One 256$\times$256 image & 0.937\tabularnewline
\hline
\multirow{3}{2cm}{Services} &Ephemeris calculation & $\sim$0.001 \tabularnewline
\cline{2-3}
 & Weather retrieval & $\sim$0.002 \tabularnewline
\cline{2-3}
 & Instrument Status retrieval & $\sim$0.001\tabularnewline
\hline
\end{tabular}
\begin{flushleft}
Note. TO -- Time Overhead (second)
\end{flushleft}
\end{table}

According to the Table~\ref{tblTimeOverhead}, it is easy to estimate the time overhead of the tasks. For example, to generate a UVFITS file with one process, it should take at least 0.72 second (The time overhead of data preprocessing + UVFITS Generation).

To further improve the performance, the multi-threaded and multi-process technologies has been used in the DDPP.
The DDPP is a multi-threaded application that permits more threads to run tasks. According to the hardware configuration of the servers, we setup the number of the threads as the number of CPU cores. At least 32 threads are started in one server to improve the processing performance. For example, if four servers in the cluster work in parallel, there are total at least 128 threads are used to parallel generate UVFITS files. Therefore, under current hardware environment, about 178 files would be generated in one second.

\section{Discussion}

The DDPP is a distributed parallel computing pipeline for CSRH. Although the construction of the DDPP has been completed in the main, there are still some issues need to be discussed and further improved.

{\em1. OpenCluster}

The OpenCluster is a new lightweight infrastructure for designing astronomy data processing pipeline. Objectively, it is risky to build the DDPP on the OpenCluster infrastructure for CSRH. Many mature traditional techniques such as Message Passing Interface (MPI)~\citep{gropp1999using}, Hadoop (http://hadoop.apache.org/), and Storm (http://storm.apache.org/) have been widely used in data processing, and have lots of successful cases. For example, according to the documents of SKA, stream computing technique would be deployed in the SKA's high performance storage and data processing. MPI technology is widely used in high performance image processing for many modern telescopes.

We ultimately develop the OpenCluster instead of these mature technologies because these mature systems provide several useful features to easily construct the high performance distributed computing programs, but it is difficult to implement CSRH data processing pipeline by using these systems.

1) The data processing of CSRH has many different and variable requirements of data reduction and data productions. For example, the data productions of CSRH would be different formats such as raw data with data pro-processing only, UVFITS, FITS-IDI, dirty image or deconvolution image. For traditional technologies especially the MPI technology, it is very difficult to process different tasks in parallel which would lead to losing the advantage of data parallel processing.

2) Traditional distributed computing infrastructures such as MPI are hard to support data driven mode which urgently demanded by CSRH.

3) These infrastructures are difficult and complex for astronomers to construct their pipeline system because the astronomers have to learn many profound theories such as process, thread,  mutex and semaphores. Meanwhile, these astronomers also need to master the programming skill on distributed computing programming.

{\em2. Advantages and Disadvantages of the DDPP}

The DDPP is the first high performance distributed astronomical data processing system in China. After continuous system tests and improvements,
the DDPP testifies the preferable improvement of the reliability and availability of the equipment with the continuous operation in the period of time. According to the feedbacks of the users, the DDPP has the following distinguished advantages.

1) Expandable. The DDPP is a loose-coupled system. All processing components are encapsulated into the standalone services and deployed upon the network. It is easier to build more service components and deployed to expand the functions of the DDPP. Meanwhile, the deadly errors in a service would not interrupt the operation of the DDPP.

2)  Robust. Referring to the current mature systems, the MQ technology is used for data and control message transferring among services. Message queues provide an asynchronous communications protocol, meaning that the sender and receiver of the message do not need to interact with the message queue at the same time. Messages placed onto the queue are stored until the recipient retrieves them. Therefore, the DDPP is a robust system that can be reliable operation without any maintenance.

3) Supporting Hybrid computing. The DDPP has integrated distributed computing technology and GPU technology for high performance data processing. Actually, a single technique is hard to meet the requirements of high performance data processing of CSRH. GPU is suitable for high performance image processing, but is hard to deal with the situations of massive data communication and transfer. Traditional parallel computing technology such as MPI has significantly disadvantages on communications between each cluster nodes. Communications would create a large overhead while processing massive data of CSRH.

However, the performance is a considerable problem of the DDPP in the present.
The using of Python language brings more advantages for the DDPP such as good scalability and portability especially the availability of many scientific computing packages. Python language is becoming a main stream computer language in current scientific data processing. There are many mature scientific data packages such as SunPy, AstroPy, NumPy would improve the development performance and guarantee the correctness of data reduction. However,
according to the results shown in Table~\ref{tblTimeOverhead}, the computing performance of each processing task which is written by Python language is not very high while comparing to the performance of C/C++ language implementation. In some computing tasks, the performance of C/C++ would be at least 2 times faster than that of Python.

In addition, the performance of image deconvolution is still a big problem for current system. Due to the limitation of the H\"ogbom CLEAN algorithm, the multiple iterations would lead to a very time consuming CLEAN.

The improvement for computing performance is one of the most significant tasks in the future. Meanwhile, with the quick decrease of the computer hardware price, the program written by Python could also meet the requirements of high performance data processing by purchasing more computers.

\section{Conclusion}
CSRH is to be an important synthetic aperture radio interferometry for obtaining high quality radio images at frequency range from 400 MHz to 15 GHz with high temporal, spatial, and spectral resolution. To meet the requirements of CSRH data processing, in the study, we have implemented and deployed the DDPP that has many distinguished features. Meanwhile, the key techniques such as raw data archive, data index creation, high precision target position calculation, and high performance FITS file generation are proposed in detail.

In summary, the successful application of the DDPP proves that the open source distributed computing infrastructure (i.e., OpenCluster developed by our own and can be download at https://github.com/astroitlab/opencluster) is robust, reliable and scalable. The distributed computing technology should be a trend for developing high performance data processing pipeline for modern telescopes.
Our study presents an valuable reference for other radio telescopes especially aperture synthesis telescopes, and also gives an valuable contribution to the current and/or future data intensive astronomical observations.

\acknowledgments

This work is supporpted by the National Natural Science Foundation of China (No. U1231205, 61462053,11263004, 11203011, 11163004 and 11103005) and Natural Science Foundation of Yunnan Province (No. 2013FA013, 2013FA032, 2013FZ018). We do appreciate the suggestions from Prof. Monique Pick of Paris Observatory and Prof. Hiroshi Nakajima of National Astronomical Observatory of Japan. The authors also gratefully acknowledge the helpful comments and suggestions of the reviewers.

\bibliography{CSRH}

\begin{thebibliography}{33}
\expandafter\ifx\csname natexlab\endcsname\relax\def\natexlab#1{#1}\fi

\bibitem[{Barrett \& Bridgman(1999)}]{barrett1999pyfits}
Barrett, P. \& Bridgman, W. 1999, in Astronomical Data Analysis Software and
  Systems VIII, Vol. 172, 483

\bibitem[{Beck \& Gaensler(2004)}]{Beck2004}
Beck, R. \& Gaensler, B. 2004, New Astronomy Reviews, 48, 1289

\bibitem[{Buck {et~al.}(2004)Buck, Foley, Horn, Sugerman, Fatahalian, Houston,
  \& Hanrahan}]{buck2004brook}
Buck, I., Foley, T., Horn, D., {et~al.} 2004, in ACM Transactions on Graphics
  (TOG), Vol.~23, ACM, 777--786

\bibitem[{Charlot {et~al.}(1995)Charlot, Sovers, Williams, \&
  Newhall}]{Charlot2005}
Charlot, P., Sovers, O., Williams, J., \& Newhall, X. 1995, The Astronomical
  Journal, 109, 418

\bibitem[{Dewdney {et~al.}(2009)Dewdney, Hall, Schilizzi, \&
  Lazio}]{Dewdney2009}
Dewdney, P.~E., Hall, P.~J., Schilizzi, R.~T., \& Lazio, T. J.~L. 2009,
  Proceedings of the IEEE, 97, 1482

\bibitem[{Freudling {et~al.}(2013)Freudling, Romaniello, Bramich, Ballester,
  Forchi, Garcia-Dablo, Moehler, \& Neeser}]{Freudling2013}
Freudling, W., Romaniello, M., Bramich, D.~M., {et~al.} 2013, ASTRONOMY \&
  ASTROPHYSICS, 559

\bibitem[{Glendenning \& Raffi(2008)}]{glendenning2008alma}
Glendenning, B. \& Raffi, G. 2008, in SPIE Astronomical Telescopes+
  Instrumentation, International Society for Optics and Photonics,
  701902--701902

\bibitem[{Greisen(2011)}]{greisen2011fits}
Greisen, E.~W. 2011, AIPS Memo Series, 114r, Socorro, New Mexico, USA

\bibitem[{Greisen(2012)}]{greisen2012aips}
Greisen, E.~W. 2012

\bibitem[{Gropp {et~al.}(1999)Gropp, Lusk, \& Skjellum}]{gropp1999using}
Gropp, W., Lusk, E., \& Skjellum, A. 1999, Using MPI: portable parallel
  programming with the message-passing interface, Vol.~1 (MIT press)

\bibitem[{Hintjens(2013)}]{hintjens2013zeromq}
Hintjens, P. 2013, ZeroMQ: Messaging for Many Applications (" O'Reilly Media,
  Inc.")

\bibitem[{H{\"o}gbom(1974)}]{hogbom1974aperture}
H{\"o}gbom, J. 1974, Astronomy and Astrophysics Supplement Series, 15, 417

\bibitem[{Hummel {et~al.}(2006)Hummel, Modigliani, Szeifert, \&
  Dumas}]{Hummel2006}
Hummel, W., Modigliani, A., Szeifert, T., \& Dumas, C. 2006, New Astronomy
  Reviews, 50, 412

\bibitem[{Jenness \& Economou(2014)}]{Jenness2014}
Jenness, T. \& Economou, F. 2014, Astronomy and Computing,

\bibitem[{Kaplan {et~al.}(2012)Kaplan, Bartlett, Monet, Bangert, Puatua,
  Harris, Fredericks, Barron, \& Barrett}]{kaplan2012novas}
Kaplan, G., Bartlett, J.~L., Monet, A., {et~al.} 2012, Astrophysics Source Code
  Library, 1, 02003

\bibitem[{Liu {et~al.}(2014)Liu, Wang, Ji, Deng, Dai, \& Liang}]{LYB2014}
Liu, Y.-b., Wang, F., Ji, K.-f., {et~al.} 2014, Journal of the Korean
  Astronomical Society, 47, 115

\bibitem[{McMullin {et~al.}(2007)McMullin, Waters, Schiebel, Young, \&
  Golap}]{mcmullin2007casa}
McMullin, J., Waters, B., Schiebel, D., Young, W., \& Golap, K. 2007, in
  Astronomical Data Analysis Software and Systems XVI, Vol. 376, 127

\bibitem[{Neumeyer {et~al.}(2010)Neumeyer, Robbins, Nair, \&
  Kesari}]{neumeyer2010s4}
Neumeyer, L., Robbins, B., Nair, A., \& Kesari, A. 2010, in Data Mining
  Workshops (ICDMW), 2010 IEEE International Conference on, IEEE, 170--177

\bibitem[{Newhall {et~al.}(1983)Newhall, Standish, \& Williams}]{Newhall1983}
Newhall, X., Standish, E.~M., \& Williams, J.~G. 1983, Astronomy and
  Astrophysics, 125, 150

\bibitem[{Sault {et~al.}(1995)Sault, Teuben, \&
  Wright}]{sault1995retrospective}
Sault, R.~J., Teuben, P.~J., \& Wright, M.~C. 1995, in Astronomical Data
  Analysis Software and Systems IV, Vol.~77, 433

\bibitem[{Shamir \& Nemiroff(2008)}]{Shamir2008}
Shamir, L. \& Nemiroff, R.~J. 2008, Applied Soft Computing, 8, 79

\bibitem[{Standish {et~al.}(1998)Standish, Planetary, \&
  Ephemerides}]{Standish1998}
Standish, E., Planetary, J., \& Ephemerides, L. 1998, JPL IOM 312. F-98, 48, 1

\bibitem[{Thompson {et~al.}(2008)Thompson, Moran, \& Swenson~Jr}]{Thompson2008}
Thompson, A.~R., Moran, J.~M., \& Swenson~Jr, G.~W. 2008, Interferometry and
  synthesis in radio astronomy (John Wiley \& Sons)

\bibitem[{Urvashi {et~al.}(2003)Urvashi, Rao, \& NCRA}]{urvashi2003automatic}
Urvashi, R., Rao, A.~P., \& NCRA, P. 2003, Automatic RFI identification and
  flagging, Tech. rep., Tech. Rep

\bibitem[{Wang {et~al.}(2011)Wang, Yan, Liu, Chen, \& Liu}]{WangWei2011}
Wang, W., Yan, Y., Liu, D., Chen, Z., \& Liu, F. 2011, URSI GASS, Istanbul,
  Turkey

\bibitem[{Wells {et~al.}(1981)Wells, Greisen, \& Harten}]{Wells1981}
Wells, D., Greisen, E., \& Harten, R. 1981, Astronomy and Astrophysics
  Supplement Series, 44, 363

\bibitem[{Wu(2005)}]{Wu2005}
Wu, K. 2005, in Journal of Physics: Conference Series, Vol.~16, IOP Publishing,
  556

\bibitem[{Wu {et~al.}(2009)Wu, Ahern, Bethel, Chen, Childs, Cormier-Michel,
  Geddes, Gu, Hagen, Hamann, {et~al.}}]{Wu2009}
Wu, K., Ahern, S., Bethel, E.~W., {et~al.} 2009, in Journal of Physics:
  Conference Series, Vol. 180, IOP Publishing, 012053

\bibitem[{Wu {et~al.}(2001)Wu, Otoo, Shoshani, \& Nordberg}]{Wu2001}
Wu, K., Otoo, E.~J., Shoshani, A., \& Nordberg, H. 2001, Lawrence Berkeley
  National Laboratory, Tech. Rep

\bibitem[{Yan {et~al.}(2010)Yan, Huang, Chen, Liu, \& Tan}]{YAN2010}
Yan, Y., Huang, J., Chen, B., Liu, Y., \& Tan, C. 2010, Advances in Space
  Research, 46, 413

\bibitem[{Yan {et~al.}(2011)Yan, Zhang, Chen, Wang, Liu, \& Geng}]{YAN2011}
Yan, Y., Zhang, J., Chen, Z., {et~al.} 2011, in General Assembly and Scientific
  Symposium, 2011 XXXth URSI (IEEE), 1--4

\bibitem[{Yan {et~al.}(2004)Yan, Zhang, \& Huang}]{YAN204}
Yan, Y., Zhang, J., \& Huang, G. 2004, in Radio Science Conference, 2004.
  Proceedings. 2004 Asia-Pacific (IEEE), 391--392

\bibitem[{Yan {et~al.}(2009)Yan, Zhang, Wang, Liu, Chen, \& Ji}]{YAN2009}
Yan, Y., Zhang, J., Wang, W., {et~al.} 2009, Earth, Moon, and Planets, 104, 97

\end{thebibliography}
\bibliographystyle{pasp}
\end{document}